\documentclass[11pt]{article}

\usepackage{fullpage}
\usepackage[T1]{fontenc}
\usepackage[usenames,dvipsnames]{color}
\usepackage{stmaryrd}

\RequirePackage[colorlinks=true]{hyperref}
\hypersetup{
  linkcolor=[rgb]{0.3,0.3,0.6},
  citecolor=[rgb]{0.2, 0.6, 0.2},
  urlcolor=[rgb]{0.6, 0.2, 0.2}
}
\usepackage{mathpazo}
\usepackage{xfrac}
\usepackage{bm}
\usepackage{todonotes}
\usepackage{setspace}
\usepackage{scrextend}
\usepackage{booktabs} 
\usepackage{enumitem}
\usepackage[american]{babel}

\usepackage{caption}
\usepackage{tikz}
\usepackage{tikz-cd}
\usetikzlibrary{snakes}
\usetikzlibrary{chains,fit}
\usetikzlibrary{shapes.geometric}
\usetikzlibrary{matrix,positioning,calc}
\usetikzlibrary{decorations.markings}
\usetikzlibrary{decorations.pathmorphing}
\tikzstyle{vertex}=[circle, draw, inner sep=0pt, minimum size=6pt]
\tikzstyle{vertbox}=[draw, inner sep=0pt, minimum size=8pt]

\usepackage[ruled]{algorithm2e}

\SetAlFnt{\small}
\SetAlCapFnt{\small}
\SetAlCapNameFnt{\small}
\SetAlCapHSkip{0pt}
\IncMargin{-\parindent}

\usepackage{rpmacros}
\usepackage{amsthm}
\usepackage{thmtools,thm-restate}

\numberwithin{equation}{section}
\declaretheoremstyle[bodyfont=\it,qed=\qedsymbol]{noproofstyle}

\declaretheorem[numberlike=equation]{observation}

\declaretheorem[name=Observation,numbered=no]{observation*}

\declaretheorem[numberlike=equation]{fact}

\declaretheorem[numberlike=equation]{problem}

\declaretheorem[numberlike=equation]{theorem}

\declaretheorem[name=Theorem,numbered=no]{theorem*}

\declaretheorem[numberlike=equation]{lemma}
\declaretheorem[name=Lemma,numbered=no]{lemma*}

\declaretheorem[name=Corollary,numbered=no]{corollary*}

\declaretheorem[name=Proposition,numbered=no]{proposition*}

\declaretheorem[name=Claim,numbered=no]{claim*}

\declaretheorem[name=Conjecture,numbered=no]{conjecture*}

\declaretheorem[name=Question,numbered=no]{question*}

\declaretheoremstyle[bodyfont=\it,qed=$\lozenge$]{defstyle} 

\declaretheorem[numberlike=equation,style=defstyle]{definition}
\declaretheorem[unnumbered,name=Definition,style=defstyle]{definition*}

\declaretheorem[unnumbered,name=Example,style=defstyle]{example*}

\declaretheorem[unnumbered,name=Notation=defstyle]{notation*}

\declaretheorem[unnumbered,name=Construction,style=defstyle]{construction*}

\declaretheorem[numberlike=equation,style=defstyle]{remark}
\declaretheorem[unnumbered,name=Remark,style=defstyle]{remark*}

\newcommand{\cost}{\mathsf{cost}}
\newcommand{\paths}{\mathcal{P}}

\newcommand{\lo}{\downarrow}
\newcommand{\up}{\uparrow}

\newcommand{\cA}{\mathcal{A}}
\newcommand{\OPT}{\mathsf{OPT}}

\newcommand{\vertex}{\node[vertex]}

\usepackage{nth}
\usepackage{intcalc}
\usepackage{xstring}

\usepackage{ifpdf}
\ifpdf
\else
\usepackage[quadpoints=false]{hypdvips}
\fi

\newcommand{\ECCCurl}[2]{http://eccc.hpi-web.de/report/\ifnumcomp{#1}{>}{93}{19}{20}#1/#2/}

\newcommand{\shortECCC}[2]{\texttt{\href{http://eccc.hpi-web.de/report/\ifnumcomp{#1}{>}{93}{19}{20}#1/#2/}{eccc:TR#1-#2}}}

\newcommand{\parseECCC}[1]{
\StrSubstitute{#1}{TR}{}[\tmpstring]%
\IfSubStr{\tmpstring}{/}{ 
\StrBefore{\tmpstring}{/}[\ecccyear]%
\StrBehind{\tmpstring}{/}[\ecccreport]%
}{
\StrBefore{\tmpstring}{-}[\ecccyear]%
\StrBehind{\tmpstring}{-}[\ecccreport]%
}%
\shortECCC{\ecccyear}{\ecccreport}}

\onehalfspace

\title{Generalized Parametric Path Problems}
\author{
	Kshitij Gajjar \thanks{School of Computing, National University of Singapore. Email: \texttt{kshitij@comp.nus.edu.sg}.}
	\and
	Girish Varma \thanks{CSTAR and ML Lab, IIIT Hyderabad, India. Email: \texttt{girish.varma@iiit.ac.in}.}
	\and
	Prerona Chatterjee \thanks{Tata Institute of Fundamental Research, Mumbai, India. Email: \texttt{prerona.ch@gmail.com}.}
	\and
	Jaikumar Radhakrishnan \thanks{Tata Institute of Fundamental Research, Mumbai, India. Email: \texttt{jaikumar@tifr.res.in}.}
}

\begin{document}
\maketitle

\begin{abstract}
	Parametric path problems arise independently in diverse domains, ranging from transportation to finance, where they are studied under various assumptions.
	We formulate a general path problem with relaxed assumptions, and describe how this formulation is applicable in these domains.
	
	We study the complexity of the general problem, and a variant of it where preprocessing is allowed. We show that when the parametric weights are linear functions, algorithms remain tractable even under our relaxed assumptions. Furthermore, we show that if the weights are allowed to be non-linear, the problem becomes $\NP$-hard. We also study the multi-dimensional version of the problem where the weight functions are parameterized by multiple parameters. We show that even with $2$ parameters, this problem is $\NP$-hard.
\end{abstract}

\section{Introduction}
	
Parametric shortest path problems arise in graphs where the cost of an edge depends on a parameter.
Many real-world problems lend themselves to such a formulation, e.g., routing in transportation networks parameterized by time/cost (\cite{C83, MS01, D04}), and financial investment and arbitrage networks (\cite{HP14, H14, M03}). Path problems have been studied independently in these domains, under specific assumptions that are relevant to the domain. For example, the time-dependent shortest path problem used to model transportation problems assumes a certain FIFO condition \autoref{eq:FIFOineq}. 
Arbitrage problems only model the rate of conversion and are defined with respect to a single currency parameter. These assumptions reduce the applicability of such algorithms to other domains.
	
We propose a generalized model for parametric path problems with relaxed assumptions, giving rise to an expressive formulation with wider applicability. 
We also present specific instances of real-world problems where such generalized models are required (see \autoref{sec:rel-works}).

\begin{definition}\label{def:gpp}
    The input to a Generalized Path Problem (GPP) is a $4$-tuple $(G,W,L,\vecx_0)$, where
    $G = (V \cup \set{s, t}, E)$ is a directed acyclic graph with two special vertices $s$ and $t$, $W =\setdef{w_e : \R^k \to \R^k}{e \in E}$ is a set of weight functions on the edges of $G$, $L \in \R^k$ is a vector used for computing the cost of a path from the $k$ parameters, and $\vecx_0 \in \R^k$ is the initial parameter.
\end{definition}
	
The aim in a GPP is to find an $s$-$t$ path $P$ in the graph $G$ that maximizes the dot product of $L$ with the \emph{composition} of the weight functions on $P$, evaluated at the initial parameter $\vecx_0$.

\begin{problem}[Generalized Path Problem (GPP)]  \label{problem:linearquery}
	\emph{Input:} An instance $(G,W,L,\vecx_0)$ of GPP.\\
	\noindent \emph{Output:} An $s$-$t$ path $P= (e_1,\ldots, e_r)$ which maximizes
	\[
	    L\cdot w_{e_r}(w_{e_{r-1}}(\cdots w_{e_{2}}(w_{e_{1}}(\vecx_0))\cdots )).
	\]
	When $k=1$, we call the GPP a \emph{scalar} GPP. Sometimes we ignore the $\vecx_0$ and just write $(G,W,L)$.
\end{problem}
		
\begin{figure*}
	\centering
	\begin{tikzpicture} [scale=0.90]
		\draw [->, gray!65, line width=1mm] (7,-1) to (14.5,-1);
		\node at (15,-1) {\large{\bf $x$}};
		\draw [->, gray!65, line width=1mm] (7,-1) to (7,5);
		\node[rotate=90] at (5.5,1.8) {\large{\bf cost of $s$-$t$ path}};
		
		\node at (5.6,4.5) {\large{\bf $P_6$}};
		\node at (5.6,-0.7) {\large{\bf $P_1$}};
		
		\draw[teal] (6.8,-0.7)--(12,4.5); \draw[step=.3, yshift=-1cm, densely dotted] (6,0) grid (6.6,0.6); \draw[very thick, red] (6,-1)--(6.6,-1)--(6.6,-0.4);
		\draw[teal] (6.8,0.2)--(14,2); \draw[step=.3, yshift=-0.1cm, densely dotted] (6,0) grid (6.6,.6); \draw[very thick, red] (6,-0.1)--(6.3,-0.1)--(6.3,0.2)--(6.6,0.2)--(6.6,0.5);
		\draw[teal] (6.8,1.32)--(14,0.6); \draw[step=.3, yshift=+0.12cm, densely dotted] (6,0.9) grid (6.6,1.5); \draw[very thick, red] (6,1.02)--(6,1.32)--(6.6,1.32)--(6.6,1.62);
		\draw[teal] (6.8,3.4)--(14,-0.2); \draw[step=.3, yshift=+0.1cm, densely dotted] (6,3) grid (6.6,3.6); \draw [very thick, red] (6,3.1)--(6.3,3.1)--(6.3,3.7)--(6.6,3.7);
		
		
		\draw[magenta, very thick] (7,-0.5)--(8,0.5)--(10,1)--(12,0.8)--(14,-0.2);
		
		\draw[teal] (6.8,2.3)--(11,4.5); \draw[step=.3, yshift=-0.1cm, densely dotted] (6,2.1) grid (6.6,2.7); \draw [very thick, red] (6,2)--(6,2.3)--(6.3,2.3)--(6.3,2.6)--(6.6,2.6);
		\draw[teal] (6.8,4.5)--(14,3); \draw[step=.3, densely dotted] (6,4.2) grid (6.6,4.8); \draw [very thick, red] (6,4.2)--(6,4.8)--(6.6,4.8);
		
		\begin{scope} [xshift=-2.4cm]
		{
			\vertex at (0,0) [fill=blue, label=below left:{\Large $s$}] (v00) {};
			\vertex at (0,2) [fill=blue] (v02) {};
			\vertex at (0,4) [fill=blue] (v04) {};
			
			\vertex at (2,0) [fill=blue] (v20) {};
			\vertex at (2,2) [fill=blue] (v22) {};
			\vertex at (2,4) [fill=blue] (v24) {};
			
			\vertex at (4,0) [fill=blue] (v40) {};
			\vertex at (4,2) [fill=blue] (v42) {};
			\vertex at (4,4) [fill=blue, label=above right:{\Large $t$}] (v44) {};
		}
		\end{scope}
		
		\begin{scope} [decoration={markings, mark=at position 0.6 with {\arrow[scale=2,>=stealth,gray]{>}}}]
			\draw [postaction={decorate}] (v00)-- node[below, opacity=0.6] {\footnotesize{$a_1x + b_1$}} (v20);
			\draw [postaction={decorate}] (v20)-- node[below, opacity=0.6] {\footnotesize{$a_2x + b_2$}}(v40);
			\draw [postaction={decorate}] (v02)-- node[above, opacity=0.6] {\footnotesize{$a_3x + b_3$}}(v22);
			\draw [postaction={decorate}] (v22)-- node[below, opacity=0.6] {\footnotesize{$a_4x + b_4$}}(v42);
			\draw [postaction={decorate}] (v04)-- node[above, opacity=0.6] {\footnotesize{$a_5x + b_5$}}(v24);
			\draw [postaction={decorate}] (v24)-- node[above, opacity=0.6] {\footnotesize{$a_6x + b_6$}}(v44);
			
			\draw [postaction={decorate}] (v00)-- node[opacity=0.6, rotate=90,yshift=0.25cm] {\footnotesize{$c_1x + d_1$}}(v02);
			\draw [postaction={decorate}] (v02)-- node[opacity=0.6, rotate=90,yshift=0.25cm] {\footnotesize{$c_2x + d_2$}}(v04);
			\draw [postaction={decorate}] (v20)-- node[opacity=0.6, rotate=90,yshift=0.25cm] {\footnotesize{$c_3x + d_3$}}(v22);
			\draw [postaction={decorate}] (v22)-- node[opacity=0.6, rotate=90,yshift=-0.25cm] {\footnotesize{$c_4x + d_4$}}(v24);
			\draw [postaction={decorate}] (v40)-- node[opacity=0.6, rotate=90,yshift=-0.25cm] {\footnotesize{$c_5x + d_5$}}(v42);
			\draw [postaction={decorate}] (v42)-- node[opacity=0.6, rotate=90,yshift=-0.25cm] {\footnotesize{$c_6x + d_6$}}(v44);
		\end{scope}
	\end{tikzpicture}
		
	\caption{
		\footnotesize{
			(Left) Graph of a GPP instance whose edge weights are linear functions of a parameter $x$. (Right) A plot of $x$ versus the costs of all possible $s$-$t$ paths $P_1,\ldots,P_6$. All 6 cost functions are linear because the composition of linear functions is linear. For example, the cost of $P_6$ is $a_6(a_5(c_2(c_1x+d_1)+d_2)+b_5)+b_6$. The table for GPP with preprocessing (PGPP) with $L=-1$ has 4 entries, indicated in pink.
		}
	}\label{fig:planar3x3}
\end{figure*}

Scalar GPP (see \autoref{fig:planar3x3}) models shortest paths by choosing weights $w_e(x)= a_e \cdot x + b_e$ and fixing $L=-1$, to convert it to a minimization problem. 
Scalar GPP also models currency arbitrage problems (\cite{HP14, CLRS09}), where the cost of a path is the product of its edge weights, by choosing weight functions to be lines passing through the origin with slopes equal to the conversion rate.

Further, GPP can model more general path problems which involve multiple parameters to be optimized. For example, in transport networks, one needs to find a path that optimizes parameters like time traveled, cost of transportation, convenience, polluting emissions, etc. (\cite{KVP20}). In finance problems, an entity can have investments in different asset classes like cash, gold, stocks, bonds, etc., and a transaction, modeled by an edge, can affect these in complex ways (see more examples in \autoref{sec:rel-works}). The edge parameters could contribute additively or multiplicatively to the cost of the path.
We study weight functions that are affine linear transformations, which allows for both of these.
	
Note that the optimal path can vary based on the value of the initial parameter $\vecx_0$. For this, we also consider a version of GPP with preprocessing (called PGPP), where we can preprocess the inputs $(G,W,L)$ and store them in a table which maps the initial values $\vecx_0$ to their optimal paths. Such a mapping is very useful in situations where the underlying network does not change too often and a large amount of computing power is available for preprocessing (e.g., the road map of a city typically does not change on a day-to-day basis). 
If the size of the table is managable, then it can be saved in memory and a query for an optimal path for a given $\vecx_0$ can be answered quickly using a simple table lookup.

\begin{problem}[GPP with Preprocessing (PGPP)] \label{problem:qgip} 
	\emph{Input:} An instance $(G,W,L)$ of GPP.\\
	\noindent \emph{Output:} A table which maps $\vecx_0$ to optimal paths.
\end{problem}

We present an efficient algorithm for scalar GPP with linear weight functions.
On the other hand, we show that if the GPP instance is non-scalar or the weight functions are non-linear, algorithms with worst-case guarantees cannot be obtained, assuming $\P\neq\NP$.

Our algorithm is based on the Bellman-Ford-Moore algorithm (\cite{B58,F56,M59}), whereas our $\NP$-hardness reductions are from two well-known $\NP$-hard problems, \textsc{Set Partition} and \textsc{Product Partition}.

The results for GPP with preprocessing (PGPP) are much more technical since they involve proving  upper and lower bounds on the number of discontinuities of the cost of the optimal path as a function of the initial parameter $x_0$.
They generalize previously known results in Time-Dependent Shortest Paths by \cite{FHS14}.
Their work crucially uses the FIFO property \autoref{eq:FIFOineq}, whereas our analysis does not make this assumption, giving a more general result.

\paragraph{Our Contributions.}
Here is a summary of our results. In \autoref{sec:rel-works}, we give specific instances from transportation and finance where these results can be applied.
\begin{enumerate}
	\item There is an efficient algorithm for scalar GPP with linear weight functions (see \autoref{sec:qgip-algo}). 
	\item Scalar PGPP with linear weight functions has a quasi-polynomial sized table, and thus table retrieval can be performed in poly-logarithmic time (see \autoref{sec:gip-upper}).
	\item  Scalar GPP with piecewise linear or quadratic weight functions is $\NP$-hard to approximate (see \autoref{sec:qgip-hard}).
	\item For scalar PGPP with piecewise linear or quadratic weight functions, the size of the table could be exponential (see \autoref{sec:gip-lower}).
	\item Non-scalar GPP (GPP with $k>1$) is $\NP$-hard (see \autoref{sec:multi-hard}).
\end{enumerate}

\section{Applications and Related Works}	\label{sec:rel-works}
	
\subsection{Applications to Transportation.}
We relate scalar GPP to an extensively well-studied problem, known as Time-Dependent Shortest Paths (TDSPs) in graphs, which  comes up in routing/planning problems in transportation networks (\cite{D04, DOS12, FHS14}).

In the TDSP setting, the parameter $x$ denotes time, and the weight $w_e(x)$ of an edge $e=(A,B)$ denotes the arrival time at $B$ if the departure time from $A$ is $x$. 
If there is another edge $e'=(B,C)$ connected to $B$, then the arrival time at $C$ along the path $(A,B,C)$ is $w_{e'}(w_e(x))$, and so on.
Thus, the cost of an $s$-$t$ path is the arrival time at $t$ as a function of the departure time from $s$. 
We say an edge $e$ is FIFO if its weight is a monotonically increasing function, i.e.,
\begin{equation} \label{eq:FIFOineq}
	x_1\leq x_2 \Longleftrightarrow w_e(x_1)\leq w_e(x_2) \qquad\forall\,x_1,x_2.
\end{equation}

The study of TDSPs can be traced back to the work of \cite{CH66}. \cite{D69} gave a polynomial time algorithm when the edges were FIFO and the queries were made in discrete time steps. These results were extended to non-FIFO networks by \cite{OR90}, and generalized further by \cite{ZM93}.

\cite{D04} summarised all known research on FIFO networks with linear edge weights. \cite{DOS12} presented an algorithm for TDSPs in this setting whose running time was at most the table size of the PGPP instance. Soon thereafter, \cite{FHS14} showed that the table size is at most $n^{O(\log n)}$, and that this is optimal, conclusively solving the problem for FIFO networks. We show that their bounds also hold for non-FIFO networks.

Some other closely related lines of work which might be of interest to the reader are \cite{MUPT21}, \cite{BGHO20}, \cite{BCV21}, \cite{RGN21}, \cite{WLT19} and \cite{D18}.

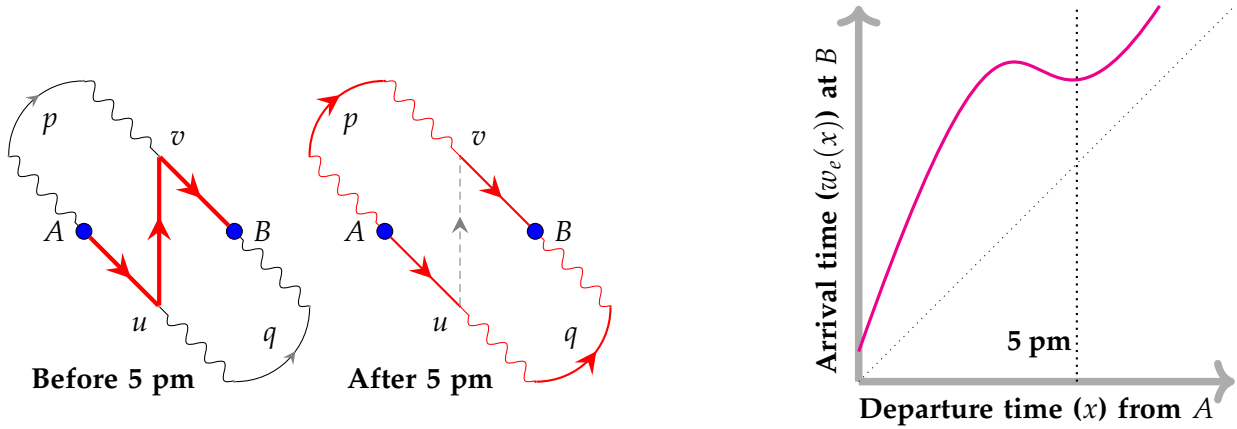
\begin{figure}
	\vspace{-6em}
	\centering
	\begin{tikzpicture}
		\begin{scope} [xshift=-2cm]
			\vertex at (0,1) [fill=blue, label=left:$A$] (v00) {};
			\vertex at (-1,2) [gray,fill=gray,minimum size=0.5pt] (v02) {};
			\vertex at (0,3) [gray,fill=gray,minimum size=0.5pt] (v20) {};
			\vertex at (1,2) [gray,fill=gray,minimum size=0.5pt, label=above right:$v$] (v22) {};
			
			\node[text width=3cm,align=left] at (.8,-1) {{\bf Before 5 pm}};

			\vertex at (-0.71,2.71) [gray,fill=gray, minimum size=0pt, label=below right:$p$] {};
			\vertex at (2.71,-0.71) [gray,fill=gray, minimum size=0pt, label=above left:$q$] {};
			
			\vertex at (1,0) [gray,fill=gray,minimum size=0.5pt, label=below left:$u$] (v11) {};
			\vertex at (2,-1) [gray,fill=gray,minimum size=0.5pt] (v14) {};
			\vertex at (3,0) [gray,fill=gray,minimum size=0.5pt] (v41) {};
			\vertex at (2,1) [fill=blue, label=right:$B$] (v44) {};
		\end{scope}
		
		\draw[snake=coil, segment aspect=0] (v02)--(v00);
		\draw[snake=coil, segment aspect=0] (v20)--(v22);
		\draw[snake=coil, segment aspect=0] (v14)--(v11);
		\draw[snake=coil, segment aspect=0] (v41)--(v44);
		
		\draw [ultra thick, red, postaction={decorate}, decoration={markings, mark=at position 0.6 with {\arrow[scale=1.5,>=stealth,red]{>}}}] (v00)--(v11);
		
		\draw [postaction={decorate}, decoration={markings, mark=at position 0.6 with {\arrow[scale=1.5,>=stealth,gray]{>}}}] (v02) to[out=90, in=180] (v20);
		
		\draw [ultra thick, red, postaction={decorate}, decoration={markings, mark=at position 0.6 with {\arrow[scale=1.5,>=stealth,red]{>}}}] (v11)--(v22);
		
		\draw [postaction={decorate}, decoration={markings, mark=at position 0.6 with {\arrow[scale=1.5,>=stealth,gray]{>}}}] (v14) to[out=0,in=270] (v41);
		
		\draw [postaction={decorate}, ultra thick, red, decoration={markings, mark=at position 0.6 with {\arrow[scale=1.5,>=stealth,red]{>}}}] (v22)--(v44);
		
		\begin{scope} [xshift=2cm]
			\vertex at (0,1) [fill=blue, label=left:$A$] (v00) {};
			\vertex at (-1,2) [red,fill=red,minimum size=0.5pt] (v02) {};
			\vertex at (0,3) [red,fill=red,minimum size=0.5pt] (v20) {};
			\vertex at (1,2) [red,fill=red, minimum size=0.5pt, label=above right:$v$] (v22) {};
			
			\node[text width=3cm,align=left] at (1,-1) {\textbf{After 5 pm}};

			\vertex at (-0.71,2.71) [red,fill=red, minimum size=0pt, label=below right:$p$] {};
			\vertex at (2.71,-0.71) [red,fill=red, minimum size=0pt, label=above left:$q$] {};
			
			\vertex at (1,0) [red,fill=red, minimum size=0.5pt, label=below left:$u$] (v11) {};
			\vertex at (2,-1) [red,fill=red,minimum size=0.5pt] (v14) {};
			\vertex at (3,0) [red,fill=red,minimum size=0.5pt] (v41) {};
			\vertex at (2,1) [fill=blue, label=right:$B$] (v44) {};
		\end{scope}
		
		\draw [red, snake=coil, segment aspect=0] (v02)--(v00);
		\draw [red, snake=coil, segment aspect=0] (v20)--(v22);
		\draw [red, snake=coil, segment aspect=0] (v14)--(v11);
		\draw [red, snake=coil, segment aspect=0] (v41)--(v44);
		
		\begin{scope} [thick, decoration={markings, mark=at position 0.6 with {\arrow[scale=2,>=stealth,red]{>}}}]
			\draw [red, postaction={decorate}] (v00)--(v11);
			\draw [red, postaction={decorate}] (v02) to[out=90,in=180] (v20);
			\draw [red, postaction={decorate}] (v14) to[out=0,in=270] (v41);
			\draw [red, postaction={decorate}] (v22)--(v44);
		\end{scope}
		
		\begin{scope} [xshift=1.8cm]
		
		\draw [densely dashed, gray, postaction={decorate}, decoration={markings, mark=at position 0.6 with {\arrow[scale=2,>=stealth,gray]{>}}}] (v11)--(v22);
		
		\draw [->, gray!65, line width=1mm] (6.5,-1) to (11.5,-1);
		
		\node[text width=5cm] at (9, -1.4) {\bf Departure time ($x$) from $A$};
		
		\draw [->, gray!65, line width=1mm] (6.5,-1) to (6.5,4);
		
		\node[text width=5cm,align=left, rotate=90] at (6.1,1.6) {\textbf{Arrival time ($w_e(x)$) at $B$}};
		
		\draw[thin, dotted] (6.5,-1)--(11.5,4);
		\draw[magenta, very thick] (6.5,-0.6) .. controls (9,6.5) and (8.5,1) .. (10.5,4);
		
		\draw[thick, dotted] (9.4,-1)--(9.4,4);
		\node at (8.9,-0.5) {{\bf 5 pm}};
		
		\end{scope}
	
	\end{tikzpicture}
	\caption{
		\footnotesize{
			An illustration of how Braess' paradox can lead to a non-FIFO edge weight function. The plot denotes $w_e(x)$ for a single edge $e=(A,B)$ of a graph. There are three routes from $A$ to $B$, namely $A$-$p$-$v$-$B$, $A$-$u$-$q$-$B$, and $A$-$u$-$v$-$B$. The roads $A$-$p$-$v$ and $u$-$q$-$B$ are quite lengthy, and thus the road linking $u$ to $v$ is preferable for a journey from $A$ to $B$. Before 5 pm, the $u$-$v$ link is available, which leads to traffic congestion on the route $A$-$u$-$v$-$B$.
			Once the $u$-$v$ link closes at 5 pm, the traffic splits equally on the routes $A$-$p$-$v$-$B$ and $A$-$u$-$q$-$B$, reducing the congestion. This leads to a drop in the travel time just around 5 pm, in accordance with Braess' paradox. During that brief interval, those departing from $A$ after 5 pm reach $B$ earlier than those departing from $A$ before 5 pm, as the plot indicates. Hence, $e$ is not FIFO (\autoref{eq:FIFOineq}).
		}
	}
	\label{fig:braess}
\end{figure}

\paragraph{Example: Braess' Paradox.} The FIFO assumption makes sense because it seems that leaving from a source at a later time might not help one reach their destination quicker. However, somewhat counter-intuitively, \cite{B68} observed that this need not always the case (\autoref{fig:braess} shows an example of Braess' paradox).
\cite{SZ83} showed that Braess' paradox can occur with a high probability. \cite{RKDG09} backed their claim with empirical evidence. In fact, there are real-world instances where shutting down a road led to a decrease in the overall traffic congestion. Two examples are Stuttgart (\cite{M70}) and Seoul (\cite[Page 71]{EK10}).
	
\subsection{Applications to Finance.} Financial domain problems have been modelled as graph problems before (\cite{DP14, KB09, E13, BAA14, A19}). We model the currency arbitrage problem (\cite{R77, SV97, DS06}) as a GPP. In the currency arbitrage problem, we need to find an optimal conversion strategy from one currency to another via other currencies, assuming that all the conversion rates are known.
	
\paragraph{Example: Multi-currency Arbitrage.} 
GPP can  model generalized multi-currency arbitrage problems. In currency arbitrage, an entity can have money available in different currencies and engage in transactions (modelled by edges) which can change the entity's wealth composition in complex ways (\cite{M03}). The transaction fees could have fixed as well as variable components, depending on the amount used. This can be modelled by affine linear transformations. Eventually the entity might liquidate all the money to a single currency, which can be modelled by the vector $L$ in the GPP instance. The goal is to pick a sequence of transactions which maximizes the cash after liquidation. Hence, this problem naturally lends itself to a GPP formulation.
	
\paragraph{Example: Investment Planning.} GPP can model investment planning by considering the nodes of the graph to be the state of the individual (which could be qualifications, contacts, experience, influence, etc). At each given state, the individual has a set of investment opportunities which are represented by directed edges. Every edge represents an investment opportunity, and the weight of the edge models the return as a function of the capital invested. Suppose an individual initially has $y$ amount of money and makes two investments in succession with returns $r_1(x), r_2(x)$, then the individual will end up with $r_2(r_1(y))$ amount of money. Though a generic investment plan could allow multiple partial investments, there are cases where this is not possible. For example, the full fees needs to be paid up front for attending a professional course or buying a house, which motivates restricting to investment plans given by paths. The vertices $s$, $t$ denote the start and end of an investment period, and the optimal investment strategy is an $s$-$t$ path which maximizes the composition of functions along the path.

\section{Algorithm for Scalar GPP with Linear Weights}\label{sec:qgip-algo}

In this section, we present our algorithm for scalar GPP with linear weight functions. Formally, we show the following.

\begin{theorem} \label{thm:QGIP_linear}
    There exists an algorithm that takes as input a scalar GPP instance $(G,W,L,x_0)$ (where $G$ has $n$ vertices and $w_e(x) = a_e \cdot x + b_e$ for every edge $e$ of $G$), and outputs an optimal $s$-$t$ path in $G$ in $O(n^3)$ running time.
 \end{theorem}
 
We use \autoref{algo:qgip} for solving GPP. 
Our algorithm is similar to the Bellman-Ford-Moore shortest path algorithm (\cite{B58,F56,M59}), where they keep track of  \emph{minimum} cost paths. The only subtlety in our case is that we need to keep track of both  \emph{minimum and maximum} cost paths with at most $k$ edges from the start vertex $s$ to every vertex $v$, as $k$ varies from $1$ to $n$. The variables $p_{\max},p_{\min}$ act as parent pointers for the maximum cost path and the minimum cost path tree rooted at $s$. $r_{\max},r_{\min}$ stores the cost of the maximum and minimum cost path. The running time of \autoref{algo:qgip} is clearly $O(n^3)$, the same as the running time of the Bellman-Ford-Moore algorithm. Its correctness follows from the following observation.

\begin{observation} 
Let $a_e$ be the coefficient of $x$ in $w_e$
\begin{itemize}
    \item If $e=(u,v)$ is the last edge on a shortest $s$-$v$ path, then its $s$-$u$ subpath is either a shortest $s$-$u$ path (if $a_e$ is positive), or a longest $s$-$u$ path (if $a_e$ is negative).
    \item If $e=(u,v)$ is the last edge on a longest $s$-$v$ path, then its $s$-$u$ subpath is either a shortest $s$-$u$ path (if $a_e$ is negative), or a longest $s$-$u$ path (if $a_e$ is positive).
\end{itemize}
\end{observation}

Then, the argument is similar to the proof of the Bellman-Ford-Moore algorithm, using the optimal substructure property. Our algorithm can also handle time constraints on the edges which can come up in transport and finance problems. For example, each investment (modelled by an edge) could have a scalar value, which denotes the time taken for it to realize. The goal is to find an optimal sequence of investments (edges) from $s$ to $t$, such that the sum of times along the path is at most some constant $T$.
We can reduce such a problem to a GPP problem with a time constraint as follows.

Replace each edge $e$ by a path of length $t_e$, where $t_e$ is the time value associated with $e$. The weight function for the first edge is simply $w_e(x)$ and for the other $t_e-1$ edges, it is the identity function. Then, \autoref{algo:qgip} can be modified so that the first for-loop stops at $T$ instead of at $n-1$.

\begin{algorithm}
    \SetAlgoLined
    For  $v \in V \setminus \{ s \}$,  $r_{\text{max}}(v) = -\infty, r_{\text{min}}(v) = \infty$\;
    $r_{\text{max}}(s) = r_{\text{min}}(s) = x$\;

    \For{$k \in [1, n-1]$}{
        \For{$e=(u,v) \in E$}{
            \eIf{ $a_e \geq 0$ }{
                \If{ $r_{\text{max}}(v) < w_e(r_{\text{max}}(u))$ }{
                    $r_{\text{max}}(v) \gets w_e(r_{\text{max}}(u))$,
                    $p_{\text{max}}(v) \gets u$\;
                }
                \If{ $r_{\text{min}}(v) > w_e(r_{\text{min}}(u))$ }{
                    $r_{\text{min}}(v) \gets w_e(r_{\text{min}}(u))$,
                    $p_{\text{min}}(v) \gets u$\;
                }
            }{
                \If{ $r_{\text{max}}(v) < w_e(r_{\text{min}}(u))$ }{
                    $r_{\text{max}}(v) \gets w_e(r_{\text{min}}(u))$,
                    $p_{\text{max}}(v) \gets u$\;
                }
                \If{ $r_{\text{min}}(v) > w_e(r_{\text{max}}(u))$ }{
                    $r_{\text{min}}(v) \gets w_e(r_{\text{max}}(u))$,
                    $p_{\text{min}}(v) \gets u$\;
                }
            }
        }
    }
    \textbf{Output}: The sequence $(t, p_{\text{max}}(t), p_{\text{max}}(p_{\text{max}}(t)), \ldots, s)$ in reverse order is the optimal path at $x$ with value $r_{\text{max}}(t)$.
    \caption{GPP with linear weight functions}\label{algo:qgip}
\end{algorithm}
	
\section{Upper Bound for Scalar PGPP with  Linear Weights}\label{sec:gip-upper}

In this section, we study scalar PGPP (linear edge weights with $L=-1$), and show that the total number of different shortest $s$-$t$ paths (for different values of $x_0\in(-\infty,\infty)$) is at most quasi-polynomial in $n$. In PGPP (\autoref{problem:qgip}), we are allowed to preprocess the graph. We compute all possible shortest $s$-$t$ paths in the graph and store them in a table of quasi-polynomial size. More precisely, if $(G,W,L)$ is a scalar GPP instance (where $G$ has $n$ vertices and $w_e(x) = a_e\cdot x+b_e$ for every edge $e$ of $G$), then we show that the number of shortest $s$-$t$ paths in $G$ is at most $n^{O(\log n)}$. (For the example in  \autoref{fig:planar3x3}, this number is 4.) Since the entries of this table can be sorted by their corresponding $x_0$ values, a table lookup can be performed using a simple binary search in $\log(n^{O(\log n)}) = O((\log n)^2)$ time. Thus, a shortest $s$-$t$ path for a queried $x_0$ can be retrieved in poly-logarithmic time.

In our proof, we will crucially use the fact that the edge weights of $G$ are of the form $w_e(x) = a_e x + b_e$. Although our result holds in more generality, it is helpful and convenient to think of the edge weights from a TDSP perspective. That is, when travelling along an edge $e=(u,v)$ of $G$, if the start time at vertex $u$ is $x$, then the arrival time at vertex $v$ is $w_e(x)$.

As the edge weights are linear and the composition of linear functions is linear, the arrival time at $t$ after starting from $s$ at time $x$ and travelling along a path $P$ is a linear function of $x$, called the cost of the path and denoted by $\cost(P)(x)$. We show that the  \emph{piecewise linear lower envelope} (denoted by $\cost_G(x)$, indicated in pink in  \autoref{fig:planar3x3}) of the cost functions of the $s$-$t$ paths of $G$ has $n^{\log n+O(1)}$ pieces. Let $p(f)$ denote the number of pieces in a piecewise linear function $f$.

\begin{theorem}\label{thm:ub_linear}
	Let $\paths$ be the set of $s$-$t$ paths in $G$. 
	Then, the cost function of the shortest $s$-$t$ path, given by $\ \cost_G(x) = \underset{P:\, P\in \paths}{\min} \cost(P)(x)$, is a piecewise linear function such that 
	\[
	    p(\cost_G(x))\leq n^{\log n+O(1)}.
	\]
\end{theorem}


Before we can prove \autoref{thm:ub_linear}, we need some elementary facts about piecewise linear functions.	
Given a set of linear functions $F$, let $F_\lo$ and $F_\up$ be defined as follows.
\[
	F_\lo(x) = \min_{f:\, f\in F} f(x) \qquad\qquad F_\up(x) = \max_{f:\, f\in F} f(x)
\]
In other words, $F_\lo$ and $F_\up$ are the piecewise linear lower and upper envelopes of $F$, respectively.

\begin{fact}[Some properties of piecewise linear functions] \label{fact:piecewisefacts}
	\begin{enumerate}[label=(\roman*)]
		\item If $F$ is a set of linear functions, then $F_\lo$ is a piecewise linear concave function and $F_\up$ is a piecewise linear convex function.
		
		\item If $f(x)$ and $g(x)$ are piecewise linear  \emph{concave} functions, then $h(x)=\min\{f(x),g(x)\}$ is a piecewise linear concave function such that $p(h)\leq p(f)+p(g)$.
		
		\item If $f(x)$ and $g(x)$ are piecewise linear functions and $g(x)$ is  \emph{monotone}, then $h(x)=f(g(x))$ is a piecewise linear function such that $p(h)\leq p(f)+p(g)$.
	\end{enumerate}
\end{fact}

\begin{proof} These facts and their proofs are inspired by (and similar to) some of the observations made by  \cite[Lemma 2.1, Lemma 2.2]{FHS14}.
	\begin{enumerate}[label=(\roman*)]
		\item Linear functions are concave (convex), and the point-wise minimum (maximum) of concave (convex) functions is concave (convex).
		\item Each piece of $h$ corresponds to a unique piece of $f$ or $g$. Since $h$ is concave, different pieces of $h$ have different slopes, corresponding to different pieces of $f$ or $g$.
		\item A break point is a point where two adjoining pieces of a piecewise linear function meet. Note that each break point of $h$ can be mapped back to a break point of $f$ or a break point of $g$. As $g$ is monotone, different break points of $h$ map to different break points of $g$. \qedhere
	\end{enumerate}
\end{proof}


We now prove the following key lemma.

\begin{lemma} \label{cl:uplo_envelope}
	Let $F$ and $G$ be two sets of linear functions, and let $H=\{f\circ g \bigm| f\in F, g\in G\}$. Then
	\begin{align}
		H_\lo(x)&=\min\{F_\lo(G_\lo(x)),F_\lo(G_\up(x))\}; \label{eq:hlo}\\
		H_\up(x)&=\max\{F_\up(G_\lo(x)),F_\up(G_\up(x))\}; \label{eq:hup}\\
		p(H_\lo)&\leq 4p(F_\lo)+2p(G_\lo)+2p(G_\up); \label{eq:hlobound}\\
		p(H_\up)&\leq 4p(F_\up)+2p(G_\lo)+2p(G_\up). \label{eq:hupbound}
	\end{align}
\end{lemma}

\begin{proof}
	We will first show \autoref{eq:hlo}. Since $F$ is the set of outer functions, it is easy to see that
	\begin{equation} \label{eq:inner_outer} 
	H_\lo(x) = \min_{g:\, g\in G} F_\lo(g(x)).
	\end{equation}
	To get \autoref{eq:hlo} from \autoref{eq:inner_outer}, we need to show that the inner function $g$ that minimizes $H_\lo$ is always either $G_\lo$ or $G_\up$. 
	Fix an $x_0\in\mathbb{R}$. We will see which $g\in G$ minimizes $F_\lo(g(x_0))$. 
	Note that for every $g\in G$, we have $G_\lo(x_0)\leq g(x_0)\leq G_\up(x_0)$.
	Thus, the input to $F_\lo$ is restricted to the interval $[G_\lo(x_0),G_\up(x_0)]$. Since $F_\lo$ is a  \emph{concave} function (\autoref{fact:piecewisefacts} (i)), it achieves its minimum at either $G_\lo(x_0)$ or at $G_\up(x_0)$ within this interval. This shows \autoref{eq:hlo}.
	
	We will now show \autoref{eq:hlobound} using \autoref{eq:hlo}. 
	Since $G_\lo$ is a concave function, it has two parts: a first part where it monotonically increases and a second part where it monotonically decreases. 
	In each part, the number of pieces in $F_\lo(G_\lo(x))$ is at most $p(F_\lo)+p(G_\lo)$ (\autoref{fact:piecewisefacts} (iii)), which gives a total of $2(p(F_\lo)+p(G_\lo))$. 
	Similarly, since $G_\up$ is a convex function, it has two parts: a first part where it monotonically decreases and a second part where it monotonically increases. 
	In each part, the number of pieces in $F_\lo(G_\up(x))$ is at most $p(F_\lo)+p(G_\up)$ (\autoref{fact:piecewisefacts} (iii)), which gives a total of $2(p(F_\lo)+p(G_\up))$. 
	Combining these using \autoref{eq:hlo} and  \autoref{fact:piecewisefacts} (ii), we obtain 
	\begin{align*}
		p(H_\lo) &\leq 2(p(F_\lo)+p(G_\lo)) + 2(p(F_\lo)+p(G_\up))\\
		&= 4p(F_\lo)+2p(G_\lo)+2p(G_\up). 
	\end{align*} 
	We skip the proof of \autoref{eq:hup} and its usage to prove \autoref{eq:hupbound} because it is along similar lines.
\end{proof}

Using this lemma, we complete the proof of \autoref{thm:ub_linear}.
\begin{proof}[Proof of  \autoref{thm:ub_linear}]
	It suffices to prove the theorem for all positive integers $n$ that are powers of $2$. Let $a,b,v$ be three vertices of $G$ and let $k$ be a power of $2$. Let $\paths_v(a,b,k)$ be the set of $a$-$b$ paths $P$ that pass through $v$ such that the $a$-$v$ subpath and the $v$-$b$ subpath of $P$ have at most $k/2$ edges each ($k$ is even number since it is a power of $2$). Let $\paths(a,b,k)$ be the set of $a$-$b$ paths that have at most $k$ edges. Note that $
	    \paths(a,b,k)=\bigcup_{v\in V}\paths_v(a,b,k).
	$
	Let $f_v(a,b,k)$ be the number of pieces in the piecewise linear lower envelope or the piecewise linear upper envelope of $\paths_v(a,b,k)$, whichever is larger. 
	Similarly, $f(a,b,k)$ is the number of pieces in the piecewise linear lower envelope or the piecewise linear upper envelope of $\paths(a,b,k)$, whichever is larger. 
	Note that every path that features in the lower envelope of $\paths(a,b,k)$ also features in the lower envelope of $\paths_v(a,b,k)$, for some $v$. 
	Thus,
	\begin{equation} \label{eq:up_quasipoly}
	    f(a,b,k)\leq \sum_{v\in V}f_v(a,b,k).
	\end{equation}
	Since $G$ has $n$ vertices, $\paths(a,b,n)$ is simply the set of all $a$-$b$ paths. And since $p(\cost_G(x))$ is the number of pieces in the piecewise linear lower envelope of these paths, $p(\cost_G(x))\leq f(a,b,n)$. 
	Thus it suffices to show that $f(a,b,n)\leq n^{\log n + O(1)}$. We will show, by induction on $k$, that $f(a,b,k)\leq (8n)^{\log k}$. 
	The base case, $f(a,b,1)\leq 1$, is trivial. Now, let $k>1$ be a power of $2$. We will now show the following recurrence.
	\begin{equation}\label{eq:up_envelope}
    	f_v(a,b,k) \leq 4 \inparen{f(a,v,k/2) + f(v,b,k/2)}
	\end{equation}
	Fix a vertex $v\in V$. By induction, $f(a,v,k/2)\leq (8n)^{\log (k/2)}$ and $f(v,b,k/2)\leq (8n)^{\log (k/2)}$.
	Note that for every path $P\in\paths_v(a,b,k)$, we have $\cost(P)(x)=\cost(P_2)(\cost(P_1)(x))$, where $P_1\in\paths(a,v,k/2)$ and $P_2\in\paths(v,b,k/2)$. 
	Thus we can invoke  \autoref{cl:uplo_envelope} with $F$, $G$ and $H$ as the set of linear (path cost) functions corresponding to the paths $\paths(v,b,k/2)$, $\paths(a,v,k/2)$ and $\paths_v(a,b,k)$, respectively. 
	Applying \autoref{eq:hlobound} and \autoref{eq:hupbound}, we get
	\[ 
	    f_v(a,b,k) \leq 4 f(v,b,k/2) + 2 f(a,v,k/2) + 2 f(a,v,k/2),
	\]
	which simplifies to \autoref{eq:up_envelope}. 
	Substituting \autoref{eq:up_envelope} in \autoref{eq:up_quasipoly}, and using the fact that $|V|=n$, we get the following.
	\begin{align*}
		f(a,b,k)&\leq 4\sum_{v\in V}\left(f(a,v,k/2)+f(v,b,k/2)\right)\\
		&\leq 4n\left((8n)^{\log (k/2)}+(8n)^{\log (k/2)}\right)\\
		=(4n)\cdot 2\cdot (8n)^{\log (k/2)}=(8n)^{\log k}.
	\end{align*}
	Thus, $f(a,b,n)\leq (8n)^{\log n}=n^{\log n + 3}$.
\end{proof}

\section{Hardness of Scalar GPP with Non-linear Weights}\label{sec:qgip-hard}

In this section, we show that it is $\NP$-hard to approximate scalar GPP, even if one of the edge weights is made piecewise linear while keeping all other edge weights linear. 

\begin{theorem} \label{thm:QGIP_piecewise_linear}
     Let $(G,W,L,x_0)$ be a GPP instance with a special edge $e^*$, where $G$ has $n$ vertices and $w_e(x) = a_e x + b_e$ for every edge $e\in E(G)\setminus \{e^*\}$, and $w_{e^*}(x)$ is piecewise linear with 2 pieces. 
     Then it is $\NP$-hard to find an $s$-$t$ path whose cost approximates the cost of the optimal $s$-$t$ path in $G$ to within a constant, both additively and multiplicatively.
\end{theorem}

Note that  \autoref{thm:QGIP_piecewise_linear} implies that  \autoref{problem:qgip} with piecewise linear edge weights is $\NP$-hard.

\begin{proof} [Proof of  \autoref{thm:QGIP_piecewise_linear}] 
    We reduce from  \textsc{Set Partition}, a well-known $\NP$-hard problem  \cite[Page 226]{GJ79}\footnote{A similar reduction can be found in  \cite[Theorem 3]{NBK06}.}.
    The  \textsc{Set Partition} problem asks if a given set of $n$ integers $A = \set{a_0, \ldots, a_{n-1}}$ can be partitioned into two subsets $A_0$ and $A_1$ such that they have the same sum.
    
    We now explain our reduction. Let $\varepsilon$ be the multiplicative approximation factor and $\delta$ be the additive approximation term. Given a  \textsc{Set Partition} instance $A = \set{a_1, \ldots, a_n}$, we multiply all its elements by the integer $\ceil{\delta+1}$. 
    Note that this new instance can be partitioned into two subsets having the same sum if and only if the original instance can. 
    Furthermore, after this modification, no subset of $A$ has sum in the range $[-\delta,\delta]$, unless that sum is zero.
    Next, we define a graph instantiated by the  \textsc{Set Partition} instance.
    
    \begin{definition} \label{def:duppathgraph}
    	$G_n$ is a directed, acyclic graph, with vertex set $\inbrace{v_0, \ldots, v_n}$.
		For every $i \in \inbrace{0, \ldots, n-1}$, there are two edges from $v_i$ to $v_{i+1}$ labelled by $f_0$ and $f_1$. 
    	The start vertex $s$ is $v_0$ and the last vertex $t$ is $v_n$ (See \autoref{fig:2path}).
    \end{definition}
    
    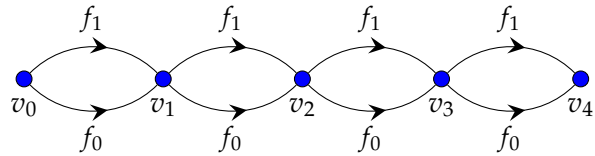
\begin{figure}[hbt!]
        \vspace{-0.7em}
        \centering
        \begin{tikzpicture}
            \foreach \i in {0,1,...,3}
            {
        	    \vertex at (1.85*\i,1) [fill=blue, label=below:$v_\i$] (v\i) {};
        	    \node at (1.85*\i+0.9,0.2) {\small $f_0$};
        	    \node at (1.85*\i+0.9,1.8) {\small $f_1$};
        	}
        	
        	\vertex at (7.4,1) [fill=blue, label=below:$v_4$] (v4) {};
        	
        	\begin{scope}[decoration={markings, mark=at position 0.6 with {\arrow[scale=2,>=stealth]{>}}}]
        	{
        	    \draw[postaction={decorate}] (v0) to[out=45,in=135] (v1);
        	    \draw[postaction={decorate}] (v0) to[out=-45,in=-135] (v1);
        	    
        	    \draw[postaction={decorate}] (v1) to[out=45,in=135] (v2);
        	    \draw[postaction={decorate}] (v1) to[out=-45,in=-135] (v2);
        	    
        	    \draw[postaction={decorate}] (v2) to[out=45,in=135] (v3);
        	    \draw[postaction={decorate}] (v2) to[out=-45,in=-135] (v3);
        	    
        	    \draw[postaction={decorate}] (v3) to[out=45,in=135] (v4);
        	    \draw[postaction={decorate}] (v3) to[out=-45,in=-135] (v4);
        	}
        	\end{scope}
        \end{tikzpicture}
        \vspace{-0.5em}
        \caption{\footnotesize{The graph $G_n$ for $n=4$.}}
        \vspace{-0.1em}
        \label{fig:2path}
    \end{figure}
    
    \begin{definition} \label{def:funpathgraph}
    	Each path of $G_n$ can be denoted by a string in $\inbrace{0,1}^n$, from left to right. For instance, if $\sigma=(0101)$, then the cost function $f_\sigma(x)$ of the path $P_\sigma$ is given by
    	\begin{equation*}
    	    f_\sigma(x) = f_{(0101)}(x) = f_1(f_0(f_1(f_0(x)))).
    	\end{equation*}
    	Note that the innermost function corresponds to the first edge on the path $P_\sigma$, and the outermost to the last.
    \end{definition}
    
    Consider the graph $G_{n+1}$. For each $i\in\inbrace{0,1,\ldots,n-1}$ and each edge $(v_i,v_{i+1})$, the edge labelled by $f_0$ has weight $x+a_i$ and the edge labelled by $f_1$ has weight $x-a_i$. Both edges from $v_n$ to $v_{n+1}$ have weight $|x|$ (and can be replaced by a single edge $e^*$). Let $\cA$ be an algorithm which solves  \autoref{problem:qgip}. 
    We will provide $G_{n+1}$ and $x_0=0$ as inputs to $\cA$, and show that $A$ can be partitioned into two subsets having the same sum if and only if $\cA$ returns a path of cost $0$.
    
    Let $\sigma = (\sigma_0 \sigma_1 \cdots \sigma_{n-1}) \in \inbrace{0,1}^n$. 
    Let $A_1$ be the subset of $A$ with characteristic vector $\sigma$, and let $A_0 = A \setminus A_1$. 
    The cost of the path $P_\sigma$ (\autoref{def:funpathgraph}) from $v_0$ to $v_n$ is 
    \[
        \cost(P_\sigma)(x) = x + \sum_{i=0}^{n-1}(-1)^{\sigma_i}a_i = x + \sum_{a_i \in A_0} a_i - \sum_{a_i \in A_1} a_i .
    \]
    Now if we set the start time from vertex $v_0$ as $x=x_0=0$, then we obtain the following. 
    \[
        \cost(P_\sigma)(0)=0 \implies \sum_{a_i \in A_0} a_i = \sum_{a_i \in A_1} a_i.
    \]
    Let $\OPT$ be a shortest path in $G_{n+1}$ and $Q$ be the path returned by $\cA$ at start time $x=x_0=0$. 
    The last edge from $v_n$ to $v_{n+1}$ (whose weight is $|x|$) ensures that $\OPT\geq 0$. So, if $\OPT=0$, then $
        \cost(Q)(0) \leq \varepsilon\cdot0 + \delta=\delta.
    $
    
    Since every path of non-zero cost in $G_{n+1}$ has cost more than $\delta$, $\cost(Q)(0)=0$ if $\OPT=0$. 
    Further, if $\OPT>0$, then $\OPT\geq\ceil{\delta+1}$, and so $\cA$ returns a path of cost more than $\delta$. 
    Thus, $A$ can be partitioned into two subsets having the same sum if and only if $\cA$ returns a path of cost $0$.
\end{proof}

\begin{remark}
	Our reduction also works if we change the weight of the last edge from $|x|$ to $x^2$, implying that scalar GPP with polynomial functions is $\NP$-hard, even if one of the edge weights is quadratic and all other edge weights are linear.
\end{remark}

\section{Lower Bound for Scalar PGPP with Non-linear Weights}\label{sec:gip-lower}

In this section, we show that for the graph $G_n$ defined in the previous section (\autoref{def:duppathgraph}) with a suitable choice of the weight functions $f_0$ and $f_1$, the table size for PGPP (\autoref{problem:qgip}) can be exponential in $n$. 
Note that $G_n$ has exactly $2^n$ paths from $s$ to $t$. 
We will show that each of these paths is a shortest $s$-$t$ path, for some value of $x$. 
Thus, there is a scalar GPP instance $(G,W,L)$ for which the table size is $2^{\Omega(n)}$, needing $\log(2^{\Omega(n)})=\Omega(n)$ time for a table lookup.

Our proof is by induction on $n$. We define the functions $f_0$ and $f_1$ in such a way that their behaviour within the interval $[0,1]$ has some very special properties, stated in  \autoref{lem:lb_main}. 
This enables us to show that the number of times the compositions of these functions achieve their minimum within the interval $[0,1]$ doubles every time $n$ increases by one.

We need some notation before we can proceed. For a function $f : \R \to \R$, and a subset $A \subseteq \R$, if $B \supseteq f(A)$, then we denote by $f|_A : A \to B,$ the function defined by $f|_A(x) = f(x)$ for every $x \in A$, also known as the restriction of $f$ to $A$.

\begin{lemma} \label{lem:lb_main}
	Suppose $f_0, f_1 : \R \to \R$ are functions such that $f_0|_{[0, 1/3]} : [0, 1/3] \to [0, 1]$ and $f_1|_{[2/3, 1]} : [2/3, 1] \to [0,1]$ are bijective.
	Further, suppose $\abs{f_0(x)} \geq 1$ for every $x \in (-\infty, 0] \cup [2/3, \infty)$; and $\abs{f_1(x)} \geq 1$ for every $x \in (-\infty, 1/3] \cup [1, \infty)$. Then, for every $n \geq 1$, there is a function $\alpha_n : \inbrace{0,1}^n \to (0,1)$ such that
	\begin{enumerate}[label=(\roman*),noitemsep,partopsep=0pt,topsep=0pt,parsep=0pt]
		\item $\alpha_n(\sigma) \in [0, 1/3]$ \, if \, $\sigma_1 = 0$;
		\item $\alpha_n(\sigma) \in [2/3, 1]$ \, if \, $\sigma_1 = 1$;
		\item For every $\sigma, \tau \in \set{0,1}^n$, $\alpha_n(\sigma) = \alpha_n (\tau) \Longleftrightarrow \sigma = \tau$;
		\item For every $\sigma, \tau \in \set{0,1}^n$, $f_\sigma(\alpha_n(\tau)) = 0 \Longleftrightarrow \sigma = \tau$.
	\end{enumerate}
\end{lemma}

\begin{proof}
	As stated earlier, we prove this lemma by induction on $n$. 
	For the  \textbf{base case} ($n=1$), we define
	\[
	    \alpha_1(0) = (f_0|_{[0, 1/3]})^{-1}(0) \text{\quad \& \quad} \alpha_1(1) = (f_1|_{[2/3, 1]})^{-1}(0).
	\]
	First we check if $\alpha_1$ is well-defined and its range lies in $(0,1)$.
	To see that $\alpha_1(0)$ and $\alpha_1(1)$ are well-defined, note that the inverses of the functions $f_0|_{[0, 1/3]}$ and $f_0|_{[0, 1/3]}$ are well-defined because they are bijective.
	To see that the range of $\alpha_1$ lies in $(0,1)$, note that for $x\in\inbrace{0,1}$, we have $\abs{f_0(x)}\geq 1$ and $\abs{f_1(x)}\geq 1$, implying that they are both non-zero.
	Thus, $0<\alpha_1(x)<1$.
	
	We now show that $\alpha_n$ satisfies (i), (ii), (iii), (iv). Since $f_0|_{[0, 1/3]}$ and $f_1|_{[2/3, 1]}$ are bijective, $\alpha_1(0)\in[0,1/3]$ and $\alpha_1(1)\in[2/3,1]$.
	Thus, $\alpha_1$ satisfies (i), (ii).
	Since these intervals are disjoint, $\alpha_1$ satisfies (iii).
	Finally, note that $f_0(\alpha_1(0)) = 0 = f_1(\alpha_1(1)).$
	Also, since $\abs{f_0(x)} \geq 1$ for every $x \in [2/3, 1]$ and $\abs{f_1(x)} \geq 1$ for every $x \in [0, 1/3]$, both $f_0(\alpha_1(1))$ and $f_1(\alpha_1(0))$ are non-zero.
	Thus, $\alpha_1$ satisfies (iv).
	This proves the base case.
	
	\paragraph{Induction step $(n>1)$:}
	Assume that $\alpha_{n-1} : \inbrace{0, 1}^{n-1} \to (0,1)$ has been defined, and that it satisfies (i), (ii), (iii), (iv). We now define $\alpha_n : \inbrace{0, 1}^n \to (0,1)$. Let $\sigma \in \inbrace{0,1}^n$ be such that $\sigma = \sigma_1 \sigma'$, where $\sigma_1 \in \inbrace{0,1}$ and $\sigma' = \sigma_2 \cdots \sigma_{n} \in \inbrace{0,1}^{n-1}$. We define $\alpha_n(\sigma)$ as follows.
	\[
	    \alpha_n(\sigma) = 
	    \begin{cases}
	        (f_0|_{[0, 1/3]})^{-1}(\alpha_{n-1}(\sigma')) \qquad\qquad \mbox{ if } \sigma_1 = 0\\
	        (f_1|_{[2/3, 1]})^{-1}(\alpha_{n-1}(\sigma')) \qquad\qquad \mbox{ if } \sigma_1 = 1\\ 
	    \end{cases}
	\]
	More concisely, 
	\begin{align}
	    \alpha_n(\sigma) = (f_{\sigma_1}|_A)^{-1}&(\alpha_{n-1}(\sigma')),\label{eq:alphan}
	\end{align}
	where $A = [0, 1/3]$ when $\sigma_1 = 0$ and $A = [2/3, 1]$ when $\sigma_1 = 1$,
	Note that $\alpha_n$ is well-defined and its range lies in $(0,1)$ for the same reasons as explained in the base case.
	We will now show that $\alpha_n$ satisfies (i), (ii), (iii), (iv).
	
    (i), (ii): By definition, $(f_0|_{[0, 1/3]})^{-1} : [0, 1] \to [0, 1/3]$ and $(f_1|_{[2/3, 1]})^{-1} : [0, 1] \to [2/3, 1]$. Thus, $\alpha_n$ satisfies (i), (ii).
	
	(iii): Suppose $\sigma = \sigma_1 \sigma' \in \set{0,1}^n$ and $\tau = \tau_1 \tau' \in \set{0,1}^n$.
	Clearly if $\sigma = \tau$, then $\alpha_n(\sigma) = \alpha_n (\tau)$.
	This shows the $\Leftarrow$ direction.
	For the $\Rightarrow$ direction, suppose $\alpha_n(\sigma) = \alpha_n (\tau)$.
	Then the only option is $\sigma_1 = \tau_1$, since otherwise one of $\alpha_n(\sigma), \alpha_n(\tau)$ would lie in the interval $[0, 1/3]$ and the other would lie in the interval $[2/3, 1]$. Thus,
	\[
	    (f_{\sigma_1}|_A)^{-1}(\alpha_{n-1}(\sigma')) = (f_{\sigma_1}|_A)^{-1}(\alpha_{n-1}(\tau')),
	\]
	where $A = [0, 1/3]$ when $\sigma_1 = 0$ and $A = [2/3, 1]$ when $\sigma_1 = 1$, 
	Since $(f_{\sigma_1}|_A)$ is bijective, this means that $\alpha_{n-1}(\sigma') = \alpha_{n-1}(\tau')$. Using part (iii) of the induction hypothesis, this implies that $\sigma' = \tau'$. 
	Thus, $\alpha_n$ satisfies (iii).
		
	(iv): Suppose $\sigma = \sigma_1 \sigma' \in \set{0,1}^n$ and $\tau = \tau_1 \tau' \in \set{0,1}^n$. Let us show the $\Leftarrow$ direction first. If $\sigma = \tau$, then
	\begin{align*}
		f_\sigma(\alpha_n&(\tau)) = f_\sigma(\alpha_n(\sigma))\\
		&=f_{\sigma'}(f_{\sigma_1}(\alpha_n(\sigma))) \qquad \text{(since $\sigma=\sigma_1\sigma'$)}\\
		&=f_{\sigma'}(f_{\sigma_1}((f_{\sigma_1}|_A)^{-1}(\alpha_{n-1}(\sigma'))))~~  \text{(using \autoref{eq:alphan})}\\
		&=f_{\sigma'}((f_{\sigma_1}\circ(f_{\sigma_1}|_A)^{-1})(\alpha_{n-1}(\sigma')))\\
		&\qquad \text{(function composition is associative)}\\
		&=f_{\sigma'}(\alpha_{n-1}(\sigma')).
	\end{align*}
	Using part (iv) of the induction hypothesis, we get $f_{\sigma'}(\alpha_{n-1}(\sigma')) = 0$, which implies that $f_\sigma(\alpha_n(\tau)) = 0$.
	This shows the $\Leftarrow$ direction.
		
	For the $\Rightarrow$ direction, suppose $f_\sigma(\alpha_n(\tau)) = 0$. 
	We have two cases: $\sigma_1 = \tau_1$ and $\sigma_1 \neq \tau_1$. 
	We will show that $\sigma=\tau$ in the first case, and that the second case is impossible. 
	If $\sigma_1=\tau_1$,
	\begin{align*}
		0 &= f_\sigma(\alpha_n(\tau)) = f_{\sigma'}(f_{\sigma_1}((f_{\tau_1}|_A)^{-1}(\alpha_{n-1}(\tau'))))\\
		& = f_{\sigma'}(f_{\sigma_1}((f_{\sigma_1}|_A)^{-1}(\alpha_{n-1}(\tau'))))
		 = f_{\sigma'}(\alpha_{n-1}(\tau')),
	\end{align*}
	Using part (iv) of the induction hypothesis, $f_{\sigma'}(\alpha_{n-1}(\tau')) = 0 \Rightarrow \sigma' = \tau'$, and thus $\sigma = \tau$. This handles the case $\sigma_1=\tau_1$. 
	We will now show by contradiction that the case $\sigma_1\neq\tau_1$ is impossible.
	
	Suppose $\sigma_1 \neq \tau_1$. 
	Let $\sigma_1 = 0$ and $\tau_1 = 1$ (the proof for $\sigma_1 = 1$ and $\tau_1 = 0$ is similar). Using the induction hypothesis, $(f_{\tau_1}|_{[2/3,1]})^{-1}(\alpha_{n-1}(\tau')) \in [2/3, 1]$. 
	Since $\abs{f_0(x)} \geq 1$ for every $x \in (-\infty, 0] \cup [2/3, \infty)$, this means that $\abs{f_{\sigma_1}((f_{\tau_1}|_{[2/3,1]})^{-1}(\alpha_{n-1}(\tau')))} \geq 1$. 
	Also note that if $\abs{x} \geq 1$, then both $\abs{f_0(x)} \geq 1$ and $\abs{f_1(x)} \geq 1$. By repeatedly applying this fact, it is easy to see that
	\begin{align*}
	    &\abs{f_\sigma(\alpha_n(\tau))}\\
	    &= \abs{f_{\sigma_n}(\cdots (f_{\sigma_1}((f_{\tau_1}|_{[2/3,1]})^{-1}(\alpha_{n-1}(\tau'))) \cdots )}\geq 1.
	\end{align*}
	We started with $f_\sigma(\alpha_n(\tau)) = 0$ and obtained $\abs{f_\sigma(\alpha_n(\tau))}\geq 1$, which is clearly a contradiction.
	This completes the proof of the $\Rightarrow$ direction, and thus $\alpha_n$ satisfies (iv).
\end{proof}

\begin{theorem} \label{thm:lb_piecewise}
	Consider the graph $G_n$. Define piecewise linear functions $f_0, f_1 : \R \to \R$ as follows (see  \autoref{fig:nequalsone}).
	\[
    	f_0(x) = 
    	\begin{cases}
    	    1 - 3x,   \mbox{ if } x \leq 1/3\\
    	    3x - 1,  \mbox{ if } x \geq 1/3 
    	\end{cases}
    	f_1(x) = 
    	\begin{cases}
        	2 - 3x,  \mbox{ if } x \leq 2/3\\
        	3x - 2,  \mbox{ if } x \geq 2/3 
    	\end{cases}
	\]
	For every $n \geq 1$ and $\sigma \in \inbrace{0,1}^n$, the cost function $f_\sigma$ of the path $P_\sigma$ is a unique piece in the lower envelope formed by the cost functions $\set{f_\sigma}_{\sigma\in\inbrace{0,1}^n}$. 
	Thus, the piecewise linear shortest path cost function has $2^n$ pieces.
\end{theorem}

\begin{proof}[Proof of \autoref{thm:lb_piecewise}]
    It is easy to check that $f_0$ and $f_1$ possess the conditions needed to invoke \autoref{lem:lb_main}.
    Thus for every $n\geq 1$, there exists a function $\alpha_n$ which satisfies properties (i), (ii) and (iii) of \autoref{lem:lb_main}.
    	
    Let $n$ be a positive integer.
    Consider the graph $G_n$ (\autoref{def:duppathgraph}). 
    Each path of $G_n$ is indexed by a binary string $\sigma\in\inbrace{0,1}^n$ and has cost function $f_\sigma$ (\autoref{def:funpathgraph}).
    Note that $f_0(x)\geq 0$, $f_1(x)\geq 0$ for all $x\in\mathbb{R}$.
    Thus $f_\sigma(x)\geq 0$ for all $\sigma\in\inbrace{0,1}^n$, $x\in\mathbb{R}$.
    	
    Let $\sigma\in\inbrace{0,1}^n$.
    Using property (iii), $f_\sigma(\alpha(\sigma))=0$, and $f_\tau(\alpha(\sigma))>0$ for every $\sigma\neq\tau\in\inbrace{0,1}^n$.
    Thus, the cost function $f_\sigma$ of the path $P_\sigma$ is a unique piece (which includes the point $\alpha_n(\sigma)$) in the lower envelope formed by the cost functions $\set{f_\sigma}_{\sigma\in\inbrace{0,1}^n}$.
\end{proof}

\begin{remark}
	The proof of \autoref{thm:lb_piecewise} works for a quadratic choice of the functions $f_0$ and $f_1$ as well. However, then the degree of the composed functions blows up exponentially, thereby making their bit complexity prohibitively large.
\end{remark}
    
\section{Hardness of Non-Scalar GPP with Linear Weights}\label{sec:multi-hard}

In this section, we show that non-scalar GPP is $\NP$-hard.

\begin{theorem} \label{thm:QGIP_2D_linear}
    Let $(G,W,L,\vecx_0)$ be a GPP instance, where $G$ has $n$ vertices and each edge $e$ of $G$ is labelled by a two dimensional vector $\vecw_e(x)$. The vertices $s,t$ are labelled by two dimensional vectors $\vecx_0, \vect_0$, respectively. Then it is $\NP$-hard to compute an optimal $s$-$t$ path in $G$.
\end{theorem}

Note that~\autoref{thm:QGIP_2D_linear} implies that~\autoref{problem:linearquery} with parameter $k=2$ is $\NP$-hard.

\begin{proof} [Proof of~\autoref{thm:QGIP_2D_linear}] 
    We reduce from \textsc{Product Partition} problem, a well-known $\NP$-hard problem (\cite{NBCK10}). 
    The problem is similar to the set partition problem, except that~\emph{products} of the elements are taken instead of their~\emph{sums}. 
    Formally, the problem asks if a given set of $n$ positive integers $A = \set{a_1, \ldots, a_n}$ can it be partitioned into two subsets $A_0$ and $A_1$ such that their product is the same.
    
    We now explain our reduction. Given a~\textsc{Product Partition} instance $A = \set{a_1, \ldots, a_n}$, consider the graph $G_{n+1}$ (\autoref{def:duppathgraph}). 
    For every $i\in\inbrace{0,1,\ldots,n-1}$, there are two edges from $v_i$ to $v_{i+1}$ labelled by matrices 
    \[
        \begin{bmatrix}
            a_i & 0\\
            0 & a^{-1}_i
        \end{bmatrix}
        \text{ and }
        \begin{bmatrix}
            a^{-1}_i & 0\\
            0 & a_i
        \end{bmatrix}.
    \]
    We label $s$ by the vector $\vecx_0=[1, 1]^{\text{T}}$ and $t$ by $\vect_0=[-1, -1]^{\text{T}}$. Let $\cA$ be an algorithm which solves~\autoref{problem:linearquery} with parameter $k=2$. 
    We will provide $G_{n+1}$ as input to $\cA$, and show that $A$ can be partitioned into two subsets having the same sum if and only if $\cA$ returns a path of cost $-2$.
    	
    Let $\sigma = (\sigma_1 \cdots \sigma_n) \in \inbrace{1, -1}^n$. 
    Let $A_1$ be the subset of $A$ with characteristic vector $\sigma$, and let $A_0 = A \setminus A_1$. The cost of the path $P_\sigma$ (\autoref{def:funpathgraph}) from $v_0$ to $v_n$ is 
    \[
        \cost(P_\sigma)
        =
        \begin{bmatrix}
            1 & 1
        \end{bmatrix} 
        \cdot
        \begin{bmatrix}
            \prod_{i=1}^{n} a^{\sigma_i}_i & 0\\
            0 & \prod_{i=1}^{n} a^{-1  \cdot \sigma_i}_i
        \end{bmatrix}
        \cdot
        \begin{bmatrix}
            -1\\
            -1
        \end{bmatrix}.
    \]
    Evaluating this, we obtain $\cost(P_\sigma) = - (a + a^{-1}$), where $a = \prod_{i=1}^{n} a^{\sigma_i}_i = \prod_{a_i \in A_0} a_i \cdot \prod_{a_i \in A_1} a^{-1}_i$. 
    Further, $a = 1 \Longleftrightarrow \prod_{a_i \in A_0} a_i = \prod_{a_i \in A_1} a_i$.
    
    By the AM-GM inequality, $a + a^{-1} > 2$, for every $a \neq 1$. Therefore $-(a + a^{-1}) < -2$ for every $a \neq 1$, and so $A$ can be partitioned into two subsets whose product is the same if and only if $\cA$ returns a path of cost $-2$.
\end{proof}
        
\section{Conclusion \& Discussion}
We study Generalized Path Problems on graphs with parametric weights.
We show that the problem is efficiently solvable when the weight functions are linear, but become intractable in general when they are piecewise linear. 

We assume that weight functions are deterministic and fully known in advance. 
Modelling probabilistic and partially known weight functions and proposing algorithms for them is a direction for future work. 
Furthermore, we have assumed that only one edge can be taken at a time, resulting in an optimization over~\emph{paths}. 
This requirement could be relaxed to study~\emph{flows} on graphs with parametric weights. 
Though there is some literature on such models in route planning algorithms (\cite{LPBM17}), results with rigorous guarantees such as the ones we have presented are challenging to obtain. 
In such cases, heuristic algorithms with empirical evaluation measures might be worth exploring.

\section*{Acknoledgements}
This work was done when the first author was a postdoctoral researcher at Technion, Israel. This project has received funding from the European Union’s Horizon 2020 research and innovation programme under grant agreement No. 682203-ERC-[Inf-Speed-Tradeoff]. 
The authors from TIFR acknowledge support of the Department of Atomic Energy, Government of India, under project number RTI4001.

\bibliographystyle{alphaurlpp}
\bibliography{biblio}

\section*{Plots} \label{sec:plotsplots}
In this section, we exhibit the plots of all the $s$-$t$ paths (\autoref{def:funpathgraph}) in the graphs $G_n$ (\autoref{def:duppathgraph}) for the piecewise linear weight functions $f_0, f_1$ (\autoref{thm:lb_piecewise}), for some values of $n$.

\begin{figure}[hb!]
\centering
	\frame{\includegraphics[width=0.9\textwidth]{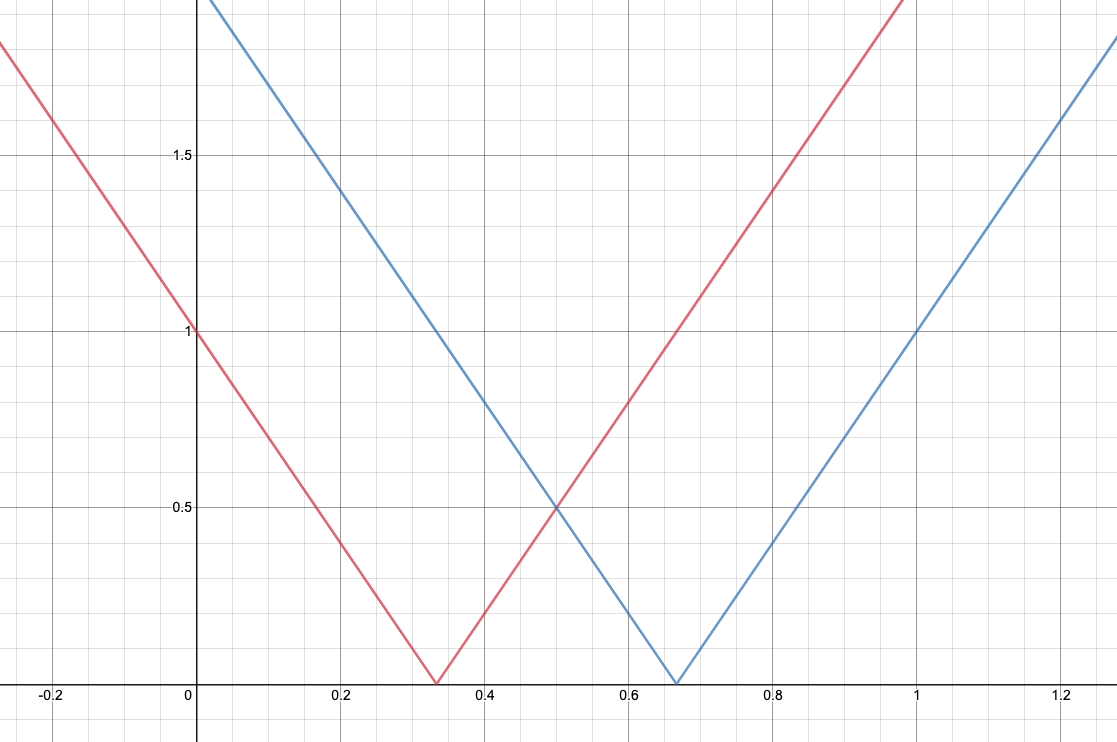}}
\caption*{\footnotesize{$n=1$}}
\end{figure} \label{fig:nequalsone}

\begin{figure}[hb!]
\centering
	\frame{\includegraphics[width=0.9\textwidth]{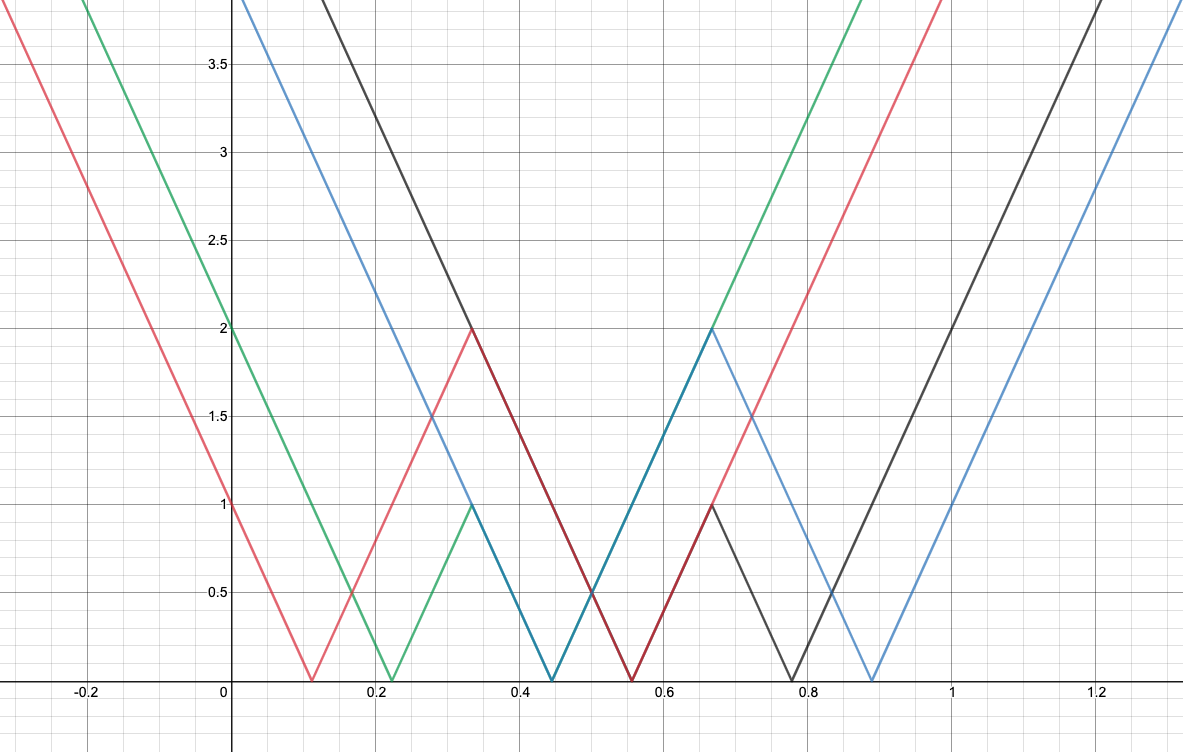}}
\caption*{\footnotesize{$n=2$}}
\end{figure}

\begin{figure}[hb!]
\centering
	\frame{\includegraphics[width=0.9\textwidth]{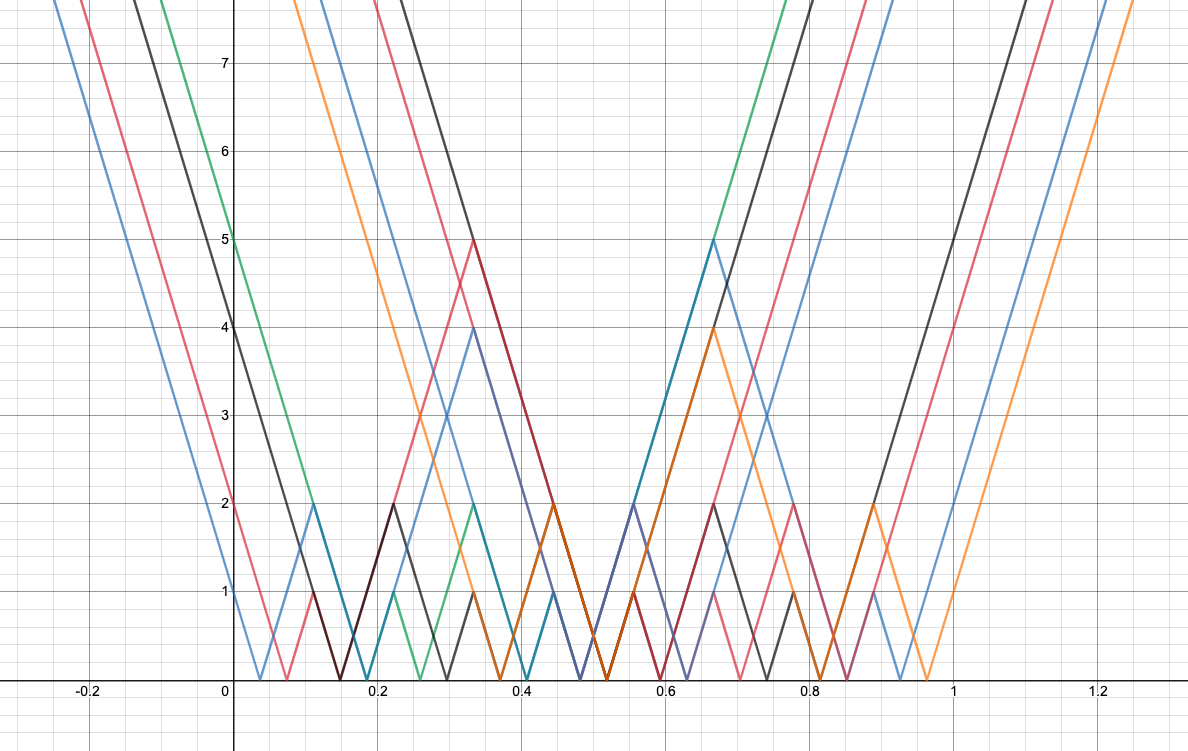}}
\caption*{\footnotesize{$n=3$}}
\end{figure}

\begin{figure}[hb!]
\centering
	\frame{\includegraphics[width=0.9\textwidth]{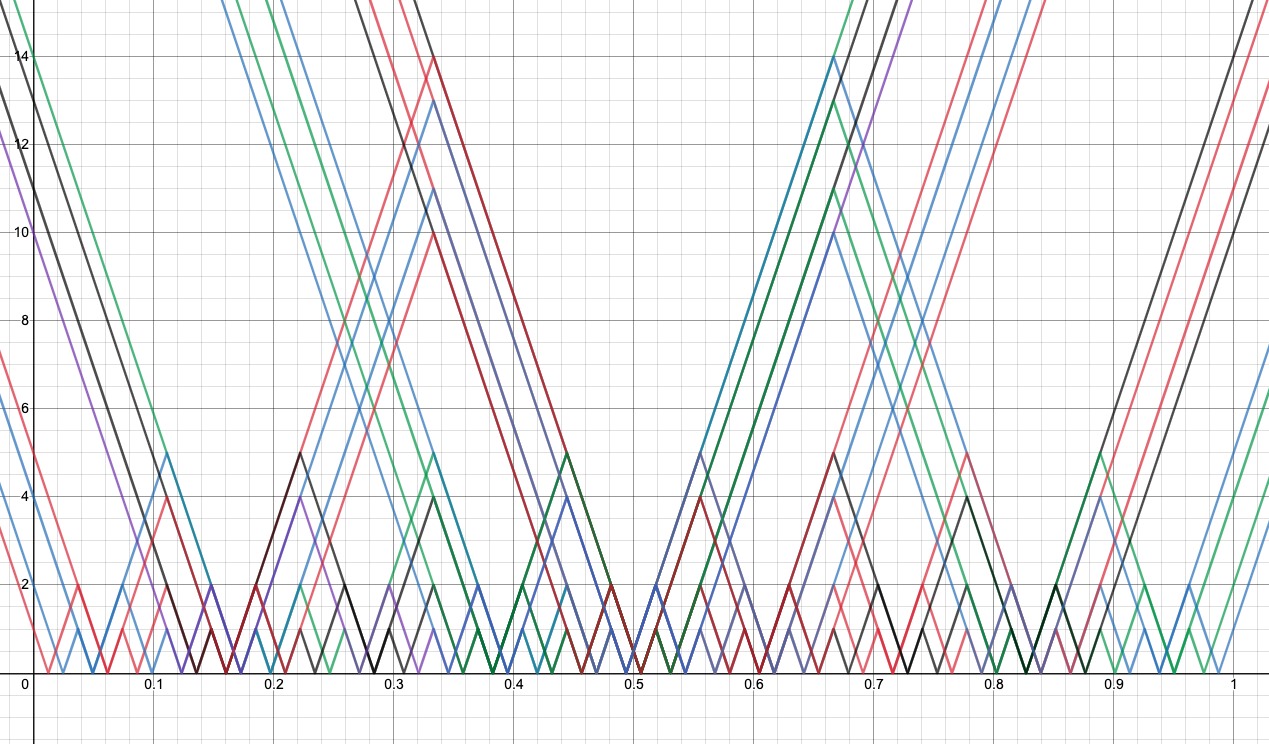}}
\caption*{\footnotesize{$n=4$}}
\end{figure}

\begin{figure}[hb!]
\centering
	\frame{\includegraphics[width=0.9\textwidth]{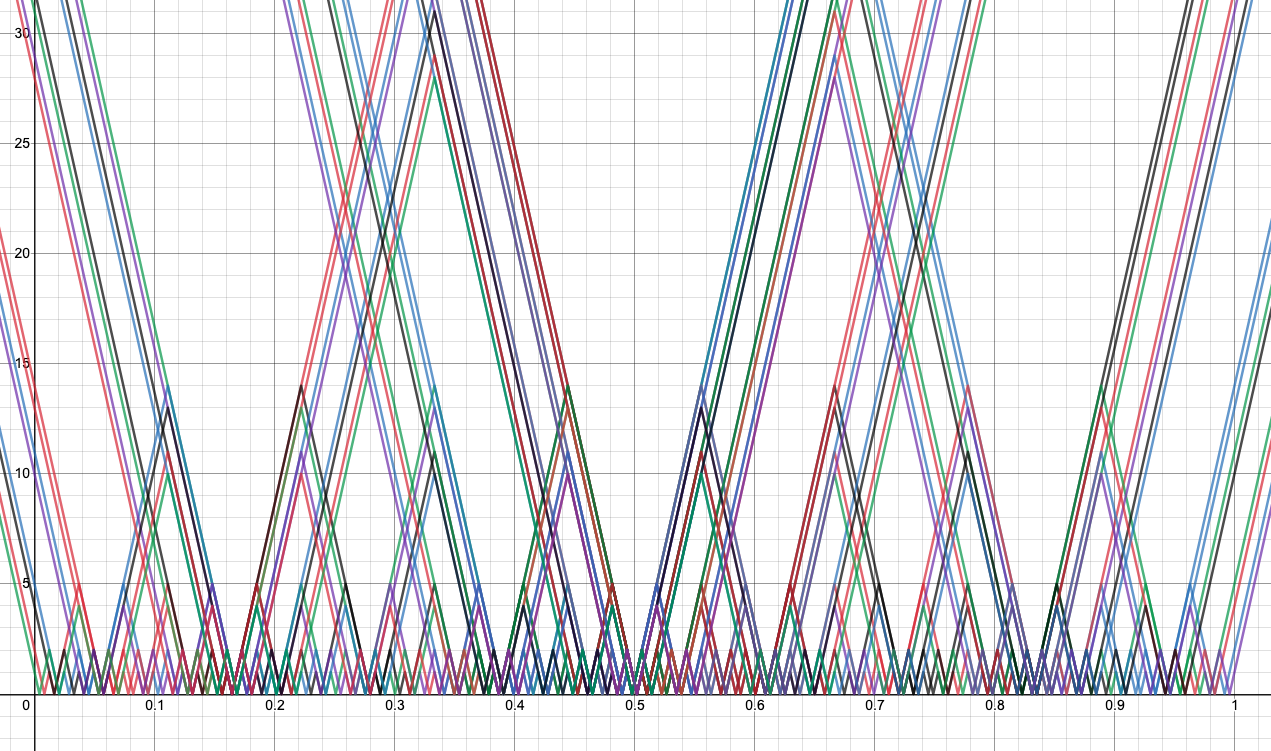}}
\caption*{\footnotesize{$n=5$}}
\end{figure}

\end{document}